\documentclass[12pt]{article}

\usepackage[english]{babel}

\usepackage{epsfig}
\usepackage{dcolumn}
\usepackage{amsmath}
\usepackage{changebar}
\usepackage{graphicx}
\usepackage{graphicx}
\usepackage{dcolumn} 
\usepackage{bm}      
\usepackage{umlaut}
\begin{document}
\renewcommand{\figurename}{{\bf Fig.}}
\renewcommand{\tablename}{{\bf Tab.}}
%
%
\title{\vspace*{-2.5cm}\hfill {\normalsize\bf BI-TP 2004/31} \\ 
\vspace*{5mm}
The influence of strange quarks on QCD phase diagram and chemical
freeze-out: \\Results from the hadron resonance gas model}
\author{A.~Tawfik\thanks{tawfik@physik.uni-bielefeld.de}\\ 
{\small University of Bielefeld, P.O.~Box 100131, D-33501~Bielefeld,
  Germany}
}
\date{
}
\maketitle

\begin{abstract} 
We confront the lattice results on QCD phase diagram for two and three
flavors with the hadron resonance gas model. Taking into account the
truncations in the Taylor-expansion of energy density $\epsilon$ done on the
lattice at finite chemical potential $\mu$, we find that the hadron
resonance gas model under the condition of constant $\epsilon$ describes
very well the lattice phase diagram. We also calculate the 
chemical freeze-out curve according to the entropy density $s$. The
$s$-values are taken from lattice QCD simulations with two and three
flavors. We find that this condition is excellent in reproducing the 
experimentally estimated parameters of the chemical freeze-out.  
\end{abstract}


\section{\label{sec:1}Introduction}

At high temperatures $T$, it is conjectured that the hadrons are dissolved into
quark-gluon plasma (QGP). 
Reducing the temperature, QGP will hadronize. At $T_{ch}$, the system goes
into chemical equilibration, freezing-out. At $T<T_{ch}$, thermal
equilibration will take place and the matter consists, once again, of
non-interacting hadrons. 

In this work, we confront the lattice results on QCD phase diagram with 
the hadron resonance gas model (HRGM)~\cite{KRT,Taw1}. We study the
influence of strange quarks on the location of the phase diagram. To
compare with two flavor lattice results, we include in HRGM the non-strange
resonances up to mass $2\;$GeV. For the three flavor results, we 
include all observed resonances with masses up to $2\;$GeV. In the
Boltzmann limit, the particle number, energy and entropy density for one
particle and its anti-particle, respectively, read   
\begin{eqnarray}
n(T,\mu) &=& \frac{g}{\pi^2} T
             m^2 K_2\left(\frac{m}{T}\right) \;
             {\mathbf \sinh\left(\frac{\mu}{T}\right)},   \label{eq:n1}  \\
\epsilon(T,\mu)  &=& \frac{g}{\pi^2} T m^2 \left[m
             K_1\left(\frac{m}{T}\right) + 3T
             K_2\left(\frac{m}{T}\right)\right] 
             \; {\mathbf \cosh\left(\frac{\mu}{T}\right)}, \label{eq:e1} \\ 
s(T,\mu) &=& \frac{g}{\pi^2}
             m^2 \left[m K_3\left(\frac{m}{T}\right)
             \; {\mathbf \cosh\left(\frac{\mu}{T}\right)} 
             - \mu K_2\left(\frac{m}{T}\right) 
             \; {\mathbf \sinh\left(\frac{\mu}{T}\right)}\right].
             \label{eq:entr} \hspace*{10mm}
\end{eqnarray}
These thermodynamic quantities will be summed over all resonances taken
into account. In mapping out the phase diagram, we take advantage of our
previous work~\cite{KRT} on simulating $T_c$ for different
quark masses and reproducing lattice thermodynamics at zero and
finite $\mu$. We assume that the critical energy density
$\epsilon_c(T_c,\mu=0)$ remains constant at all $\mu$-values. 
We suppose that this assumption is not affected by the existence of
different transitions (Tab.~\ref{Tab:1}). As done on the lattice, we can stop
the Taylor-expansion of trigonometric function in Eq.~(\ref{eq:e1}) up
to certain terms. Taking into account the quantum statistics, the results
are drawn in Fig.~\ref{Fig:2b} and Fig.~\ref{Fig:3b}.

\begin{table}[tb]
\begin{center}
\begin{tabular}{| l | c | l |}\hline %
$n_f$ & $T_c$ [MeV] & Phase transition \\  \hline\hline
$0$     & 270 & first-order  \\ \hline
$2$     & 174 & second-order or crossover \\ \hline
$3$     & 154 & first-order  \\ \hline
$2+1$   & 174 & first-order or crossover \\ \hline
\end{tabular}
\caption{\footnotesize QCD phase transitions from lattice QCD
  simulations with different flavors at $\mu=0$.}
\label{Tab:1}
\end{center}
\vspace*{-8mm}
\end{table}

The lattice results at $\mu=0$ are summarized in Tab.~\ref{Tab:1}. We
find that the strange quarks, due to the heavy mass, have small effect on
the critical values. At $\mu\neq0$, lattice simulations still suffer from
the sign-problem. The fermion determinant gets complex and therefore
MC~techniques are no longer applicable. However, considerable
progress has been made to overcome this problem~\cite{FK2,dePh1,BiSw1,DL1}.

For the chemical freeze-out curve, we compare its experimentally estimated
parameters, $T$ and $\mu_b$, with HRGM results. We propose that the entropy
density $s$ is the thermodynamic condition which drives the chemical
freeze-out. Its value is taken from lattice calculations for different
flavors at $\mu=0$. The resulting curve is seen in
Fig.~\ref{Fig:s}. Apparently, this condition gives excellent agreement with
the experimental data and reproduces very well the two characterizing
endpoints of the freeze-out curve, \hbox{($T=0$,$\mu\neq0$)} and 
\hbox{($T\neq0$,$\mu=0$)}.

\section{\label{sec:2}The QCD phase diagram}

In lattice calculations, the dependence of $T_c$ on $\mu$ can be determined 
from the pseudo-critical coupling $\beta_c(\mu)$. $T_c$ is the value at
which the susceptibility becomes maximum. From the first 
non-trivial Taylor coefficients of $T_c(\mu)$   
\begin{eqnarray} 
\frac{d^2}{d\mu^2}T_c(\mu) &=& -\frac{N_{\tau}^{-2}}{T_c(\mu=0)}
  \frac{\partial^2 \beta_c(\mu)}{\partial \mu^2}
  \left(a^{-1}\frac{\partial a}{\partial \beta}\right). \label{eq:latticeTc}
\end{eqnarray}  
$N_{\tau}$ and $a$ are the temporal lattice dimension and the lattice
spacing, respectively. The lattice beta function $\beta(a)$ is needed to
get the physical units. The results for two ~\cite{BiSw1} and three
flavors~\cite{Karsch2003}, respectively, are  
\begin{eqnarray} 
\frac{T_c(\mu_q)}{T_c(\mu_q=0)} &=& 1 -
                0.070(35)\left(\frac{\mu_q}{T_c(\mu_q=0)}\right)^2,
                \label{eq:BiSw_Tc1} \\
           &=& 1 - 0.114(46)\left(\frac{\mu_q}{T_c(\mu_q=0)}\right)^2, 
              \label{eq:BiSw_Tc3} 
\end{eqnarray}  
where $\mu_q=\mu_b/3$ is the quark chemical potential. $\mu_b$ is the
baryo-chemical potential.

\begin{figure}[t]
\begin{minipage}[t]{7.4cm}
\centerline{\includegraphics[width=7.cm]{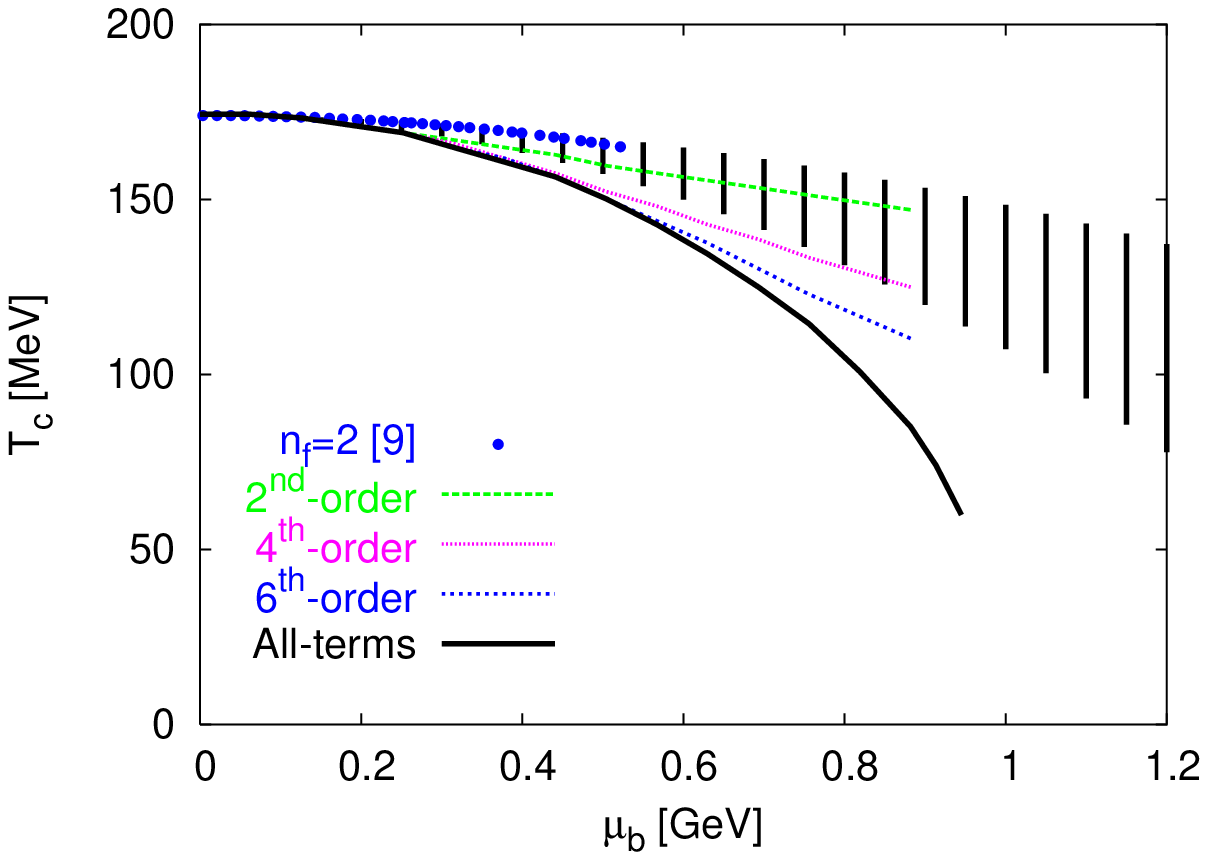}}
\caption[4.cm]{\label{Fig:2b}\footnotesize $T_c$ vs. $\mu_b$
  for light quarks. We compare lattice with HRGM results. The
  truncation up to the second-order is obviously able to describe both lattice
  simulations~\cite{BiSw1,dePh2}. The solid line gives our predictions for
  the phase diagram, when the $\epsilon_c$-expansion is not truncated.
} 
\end{minipage}\hfill
\begin{minipage}[t]{7.4cm}
\centerline{\includegraphics[width=7.cm]{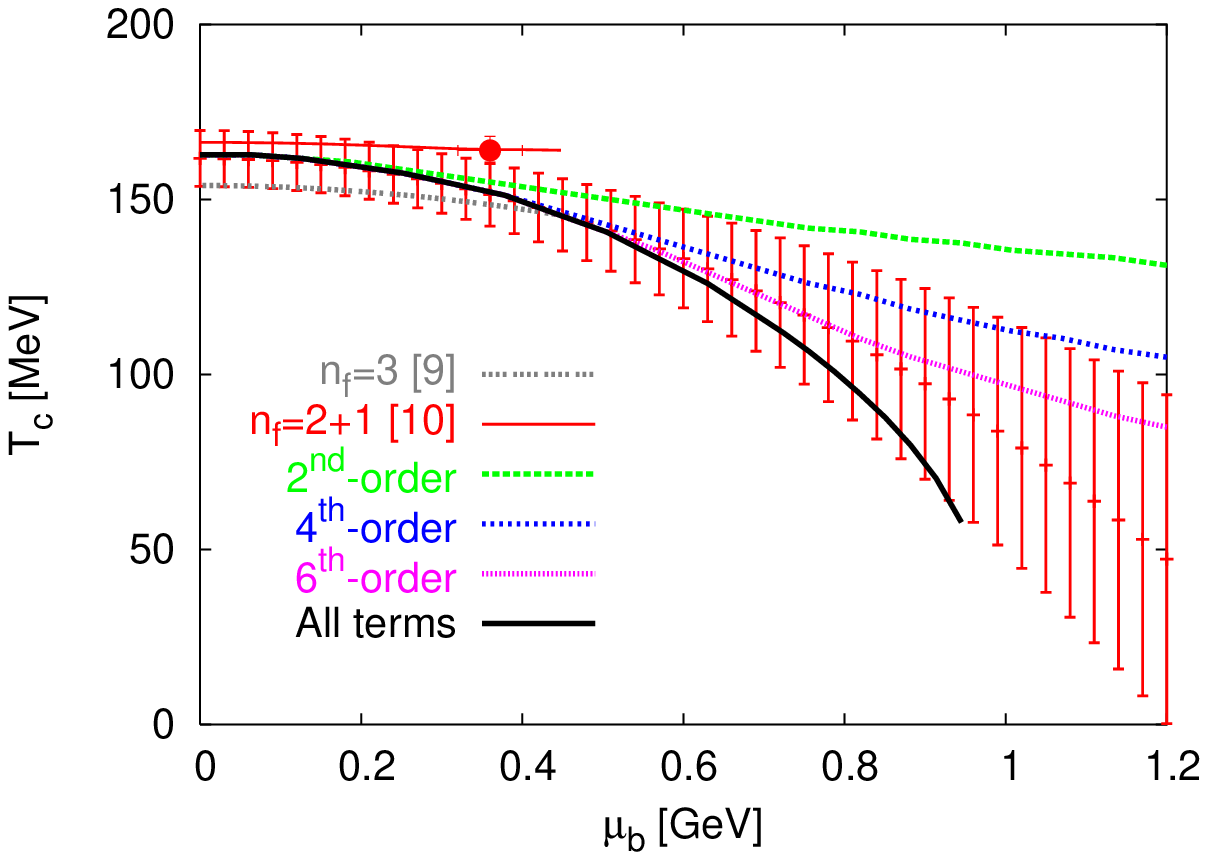}}
\caption[4.cm]{\label{Fig:3b}\footnotesize The same as in
  Fig.~\ref{Fig:2b}. Here, we include the strange quarks. Up to second-order
  truncations of $\epsilon_c$, Eq.~(\ref{eq:e1}) gives a good description
  for the lattice results~\cite{BiSw1,dePh2,FK1}. The region of most 
  reliable lattice 
  results is \hbox{$0\leq\mu_b<0.5\;$GeV}. The solid circle is the critical
  endpoint~\cite{FK1}.   
}
\end{minipage}
\vspace*{-3mm}
\end{figure}

In Fig.~\ref{Fig:2b}, we plot Eq.~(\ref{eq:BiSw_Tc1}) as vertical lines. To
compare with two flavor lattice results, we entirely exclude the strange hadron
resonances. The curves give the results obtained from different truncations
in the Taylor-expansion of $\epsilon_c$, Eq.~(\ref{eq:e1}).   
The three flavor results are shown in Fig.~\ref{Fig:3b}. Here, we
explicitly set the strange quark chemical potential $\mu_s=0$, as the case
in the lattice calculations. An extensive discussion about the dependence of
$\mu_s$ on $T$ and  $\mu_b$ is given in
Ref.~\cite{Taw1}. 

We also show the latest lattice calculations of the critical endpoint. 
According to Ref.~\cite{FK1}, it is located at
\hbox{$\mu_b^{ep}=360\pm40\;$MeV} and
\hbox{$T_c^{ep}=162\pm2\;$MeV}. In Ref.~\cite{BiSw2}, it has been found that
\hbox{$\mu_b^{ep}=420\;$MeV}. The $T_c^{ep}$-coordinate has not yet been  
calculated. We assume that the critical endpoint, at
which the phase transition is second-order, does not affect our results.

\section{\label{sec:3}The chemical freeze-out}

As discussed in Ref.~\cite{Taw2}, the entropy density is the thermodynamic
condition which guarantees chemical equilibration between reactants and
products. To prove this, we recall the theoretical chemistry. Without
energy input the chemical reactions always proceed toward    
equilibrium. The equilibrium constant ${\cal C}$ is related to the energy
difference between reactants and products $\delta\epsilon$ via 
Boltzmann factor, \hbox{${\cal C}\approx\exp(\delta\epsilon/T)$}. From 
second law of thermodynamics, \hbox{$\delta\epsilon=\delta 
  F+T\delta s +\mu\delta n$}. $\delta F$ is the free energy difference
between reactants and products. It represents the total work
in the reacting system. At equilibrium, $\delta F=0$ and, therefore,
\begin{eqnarray}
{\cal C} &\approx& \exp\left[-s - n\; \left(\frac{\mu}{T}\right)\right].
\label{eq:ch1}
\end{eqnarray}
The entropy gives the amount of energy which can't be used to produce further
work. Then, the equilibrium entropy reads
\begin{eqnarray}
s(T,\mu) &\approx& \ln\left(\frac{1}{\cal C}\right) -n(T,\mu) \;
\left(\frac{\mu}{T}\right). 
\label{eq:ch2}
\end{eqnarray}

In literature, there are different models~\cite{CR99,BS02,MS03} for the
freeze-out conditions. Cleymans and Redlich~\cite{CR99} assumed
that ${\cal C}$ depends on some average energy per hadron particle. This
apparently ignores the change in the baryon number with increasing
$\mu$. Braun-Munzinger and Stachel~\cite{BS02} rectified this and assumed
that the freeze-out is given by a constant baryon number density (compare
this with Eq.~(\ref{eq:ch1})). The two models in Ref.~\cite{BS02,MS03} give
identical results. As we show in Ref.~\cite{Taw2}, there are some
discrepancies in reproducing the experimental data by all these models.

We calculate the freeze-out curve according to Eq.~(\ref{eq:entr}). In this
context, the entropy can be seen as a measure for the degree of sharing and
spreading the energy inside the system. The way of distributing the energy
is not just an average value. But the method that controls the chemical
equilibration. i.e., produces no additional work. This is the
equilibrium entropy, Eq.~(\ref{eq:ch2}).

As $T\rightarrow0$ and $\mu\neq0$, it is assumed that the system is
just a degenerate Fermi gas of nucleons. Then from Eq.~(\ref{eq:entr}), $s=0$. 
At $\mu=0$ and $T\neq0$, the system becomes degenerate Bose gas of
pions and rho-mesons. Therefore, $s$ gets a finite-value. In the intermediate
region, along the freeze-out curve, $s$ ranges between these two extreme
limits. 

\begin{figure}[t] \vspace*{2mm}
\centerline{\includegraphics[width=8cm]{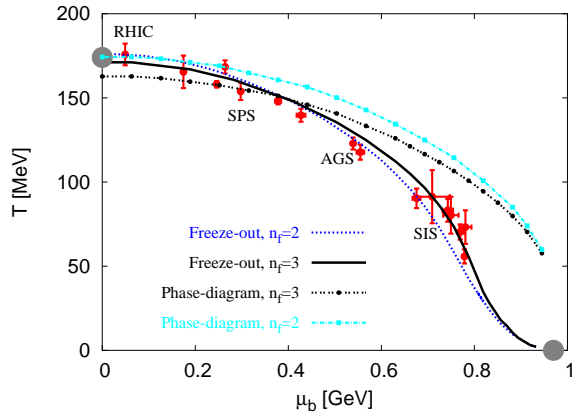}}
\caption[9.cm]{\label{Fig:s}\footnotesize
  The freeze-out curve according to the entropy density,
  Eq.~(\ref{eq:entr}). The two endpoints (large solid circles) and the
  experimentally estimated results (squares) are very well
  reproduced. There is no difference between strange and non-strange hadron
  resonances. The QCD phase diagrams for two and three quark flavors $n_f$
  are also drawn. }  
\vspace*{-3mm}
\end{figure}

At small $\mu$, it is supposed that the freeze-out and phase transition
are coincident. We can therefore use the lattice 
calculations~\cite{Taw1,Karsch2000,KRT}; $s/T^3=5$ for $n_f=2$
and $s/T^3=7$ for $n_f=3$. The normalization with respect to $T^3$ should
not be connected with the massless ideal gas. Either the hadrons in HRGM or
quarks on lattice are massive. The constant ratio $s/T^3$ obviously
fulfills Eq.~(\ref{eq:ch2}) and simultaneously obeys the third law of
thermodynamics. The quantum 
entropy~\cite{MT} is not included here. On the other hand, assignment
$s/T^3$ to a constant value and numerically solving Eq.~(\ref{eq:entr}) with
respect to $T$, leads to the  excellent results seen in
Fig.~\ref{Fig:s}. The two characterizing  endpoints and experimental data
are very well reproduced. There is almost no difference between including
and excluding strange resonances, as long as we use the corresponding
$s$-value.

\section{\label{sec:4}Conclusion}

We compared between the lattice results on the QCD phase diagram for different
flavors and the HRGM results. Taking into account the truncations in the
Taylor-expansion of the energy density $\epsilon$ done on the lattice, we
found the condition of constant energy density results in an excellent
agreement with all available lattice results. We conclude that the
influence of the strange quarks on the location of QCD phase diagram is
relatively large at small baryo-chemical potential $\mu_b$. At large
$\mu_b$, there is almost no influence. This conclusion is valid unter the
assumption, that the strange quark chemical potential $\mu_s$ is vanishing.

For mapping out the chemical freeze-out curve, we used the entropy
density. Taking its value from lattice simulations with two and three quark
flavors at $\mu=0$ and assuming it remains constant on the entire
$\mu$-axis, we obtain the meaningful results shown in Fig.~\ref{Fig:s}. The
experimentally estimated data is seen to be very well described under this
condition. The two characterizing endpoints of the chemical freeze-out
curve are also reproduced.


\begin{thebibliography}{99}

\bibitem{KRT}F.~Karsch, K.~Redlich and A.~Tawfik, Phys.~Lett.~B~{\bf
                  571}  67 (2003), Eur.~Phys.~J.~C~{\bf 29} 549 (2003), \\
         K.~Redlich, F.~Karsch and A.~Tawfik, J.~Phys.~G~{\bf
                  30}~S1271~(2004).  

\bibitem{Taw1}A.~Tawfik, {\it QCD phase diagram: comparison between lattice and
    hadron resonance gas model calculations}, in progress 

\bibitem{FK2}S.~Fodor and S.D.~Katz, Phys.~Lett.~B~{\bf 534} 87 (2002).

\bibitem{dePh1}P.~de~Forcrand and O.~Philipsen, Nucl.~Phys.~B {\bf 642}
  290 (2002). 

\bibitem{BiSw1}C.R.~Allton, {\it et.al.}, 
  Phys.~Rev.~D~{\bf 66}~074507~(2002). 

\bibitem{DL1}M.~D'Elia and M.-P.~Lombardo, Phys.~Rev.~D {\bf 67} 014505 (2003).

\bibitem{Karsch2000}F.~Karsch, E.~Laermann and A.~Peikert, Nucl.~ Phys.~B
  {\bf 605} 579 (2001).

\bibitem{Karsch2003}F.~Karsch, {\it et.al.}, 
  Nucl.~Phys.~Proc.~Suppl.~{\bf 129}~614~(2004). 

\bibitem{dePh2}P.~de~Forcrand and O.~Philipsen, Nucl.~Phys.~B {\bf 673}
  170 (2003).

\bibitem{FK1}S.~Fodor and S.D.~Katz, JHEP,~{\bf
    0404}~050~(2004).

\bibitem{BiSw2}C.R.~Allton, {\it et.al.}, 
  Phys.~Rev.~D~{\bf 68} 014507 (2003).

\bibitem{Taw2}A.~Tawfik, {\it On the conditions driving the chemical
    freeze-out},  in progress 

\bibitem{CR99}J.~Cleymans and K.~Redlich, Phys.~Rev~C,~{\bf
    60}~054908~(1999)

  
\bibitem{BS02}P.~Braun-Munzinger and J.~Stachel, J.~Phys.~G~{\bf
    28}~1971~(2002)

\bibitem{MS03}V.~Magas and H.~Satz, Eur.~J.~Phys.~C~{\bf 32}~115~(2003).


\bibitem{MT}D.~Miller and A.~Tawfik, J.~Phys.~G~{\bf 30}~731~(2004); \\
  Acta~Phys.~Polon.~B~{\bf 35}~2165~(2004),  
  hep-ph/0308192; hep-ph/0309139;  hep-ph/0312368; S.~Hamieh and A.~Tawfik,
  hep-ph/0404246.  

\end{thebibliography}
\end{document}